\documentclass[sigconf]{acmart}
\setcopyright{rightsretained} 
\def\BibTeX{{\rm B\kern-.05em{\sc i\kern-.025em b}\kern-.08emT\kern-.1667em\lower.7ex\hbox{E}\kern-.125emX}}
\usepackage[caption]{subfig}
\usepackage{float}
\usepackage{amsmath,amssymb,array,threeparttable,multirow}
\usepackage{graphicx}
\usepackage{epsfig,endnotes}
\usepackage{textcomp}
\usepackage[noend]{algpseudocode}
\usepackage[lettersize]{lscape}
\usepackage{url}
\usepackage[shortlabels]{enumitem}
    \setlist[enumerate, 1]{1.}
    \setlist{leftmargin=5.0mm}
\usepackage{tikz}
\usepackage{epstopdf}
\usepackage{adjustbox}
\usepackage{balance}
\usepackage{algorithm,algpseudocode}
\usepackage{multirow}
\usepackage{hyperref}

\algnewcommand{\algorithmicforeach}{\textbf{for each}}
\algdef{SE}[FOR]{ForEach}{EndForEach}[1]
  {\algorithmicforeach\ #1\ \algorithmicdo}
  {\algorithmicend\ \algorithmicforeach}

\begin{document}

\copyrightyear{2019} 
\acmYear{2019} 
\acmConference[MobiSys '19]{The 17th Annual International Conference on Mobile Systems, Applications, and Services}{June 17--21, 2019}{Seoul, Republic of Korea}
\acmBooktitle{The 17th Annual International Conference on Mobile Systems, Applications, and Services (MobiSys '19), June 17--21, 2019, Seoul, Republic of Korea}\acmDOI{10.1145/3307334.3328567}
\acmISBN{978-1-4503-6661-8/19/06}
\fancyhead{}

\title{
Demo: Light-Weight Programming Language for Blockchain
}

\author{Junhui Kim}
\affiliation{
    \institution{Chung-Ang University}
    \city{Seoul}
    \country{Republic of Korea}
}
\email{jhkim.cau@gmail.com}

\author{Joongheon Kim}
\affiliation{
    \institution{Chung-Ang University}
    \city{Seoul}
    \country{Republic of Korea}
}
\email{joongheon@gmail.com}

\begin{abstract}
This demo abstract introduces a new light-weight programming language \texttt{koa} which is suitable for blockchain system design and implementation. In this abstract, the basic features of \texttt{koa} are introduced including working system (with playground), architecture, and virtual machine operations. Rum-time execution of software implemented by \texttt{koa} will be presented during the session.
\end{abstract}

%
%

\ccsdesc[500]{Software and its engineering~Extensible languages}
\ccsdesc[300]{Software and its engineering~Virtual machines}
\ccsdesc[300]{Software and its engineering~Translator writing systems and compiler generators}
\ccsdesc[300]{Software and its engineering~Runtime environments}

\keywords{Blockchain; Programming Language; Smart Contract}

\maketitle

\section{Introduction}
Recently, blockchain technologies are getting a lot of attentions due to its decentralized scalable nature. Among them, Bitcoin and Ethereum are major leading blockchain technologies. 
The Bitcoin is a blockchain-based cryptocurrency system which was initiated by Nakamoto Satoshi in 2008~\cite{asiaccs2018kim,secon2018kim}. 
The trading in Bitcoin conducts via peer-to-peer networking under the encryption/description with SHA-256.
All transactions are stored as a data structure called Merkle tree by every node participating in blockchain.
The Bitcoin solves the double-spending problem using Proof-of-Work (PoW) consensus algorithm for generating blocks.
The Ethereum is a distributed computing platform based on blockchain developed by Vitalik Buterin in 2015.
Similar to the Bitcoin, All transactions are stored by every node in Etheruem.
However, it can execute applications on the blockchain using \textit{Smart Contract} as well as \textit{trading cryptocurrency}.
Thus, Etheruem applies the blockchain to various areas such as SNS, e-mail, games, and etc.

In this paper, we propose a new programming language which is easy to understand and write a code that implements the blockchain system in a scalable way. 
This language, called \texttt{Koa}, solves difficulties of understanding and analyzes the code what the script language of the Bitcoin has with high-level language style.
Since the script language is a sequence of instructions, it is difficult to understand the flow of codes at a glance.
In addition, it is tricky to implement complicated logics as instructions.
For example, when expressing a branch of conditional statement in script language, there is no distinction between the parts to compare a condition, to execute \texttt{if-true}, and to execute \texttt{if-false}.
Besides, it is not easy to figure out what each instruction is used for.
The verifying process is performed by sequentially running the unlocking and lokcing script.
It is difficult to understand what kind of actions to perform because this script code is not distinguishable by semantic units.
Furthermore, when scripts are implemented, there is a difficulty to program considering the status of stack.
If the function is defined in \texttt{koa}, it can be distinguishable by semantic units.
When verifying a value, it needs to define a new instruction or to assemble many instructions complicatedly.
On the other hand, \texttt{Koa} can implement this by combining supported operators.
Using \texttt{Koa}, it has better readability than script languages and is easier to write the code simply.

Meanwhile, \textit{solidity} is used as a smart contract language for Ethereum.
If a transaction is executed and the state changes, the Ethereum uses \textit{gas}.
Applying \textit{gas} as a kind of transaction fees, Ethereum prevents infinite loop attack.
However, the \textit{gas} makes solidity code more complicated and hinder usability.
Furthermore, the state of Ethereum makes the static analysis of solidity code difficult.
Since the result of executing function differs according to state, it is difficult to predict \textit{gas} consumption.
However, because the \texttt{Koa} is a stateless language, it can be statically analyzed.
Since there are not any branches of code execution depending on state nor loop, static analysis for the code is easy.
In addition, it can prevent the infinite loop attack.
This allows the \texttt{Koa} to predict the costs associated with running the smart contract.
Therefore, the \texttt{Koa} which is high-level language helps programmers to implement the blockchain quickly and easily.

\section{System Overview}

\begin{figure}[t]
    \centering
        \includegraphics[width =1.0\linewidth]{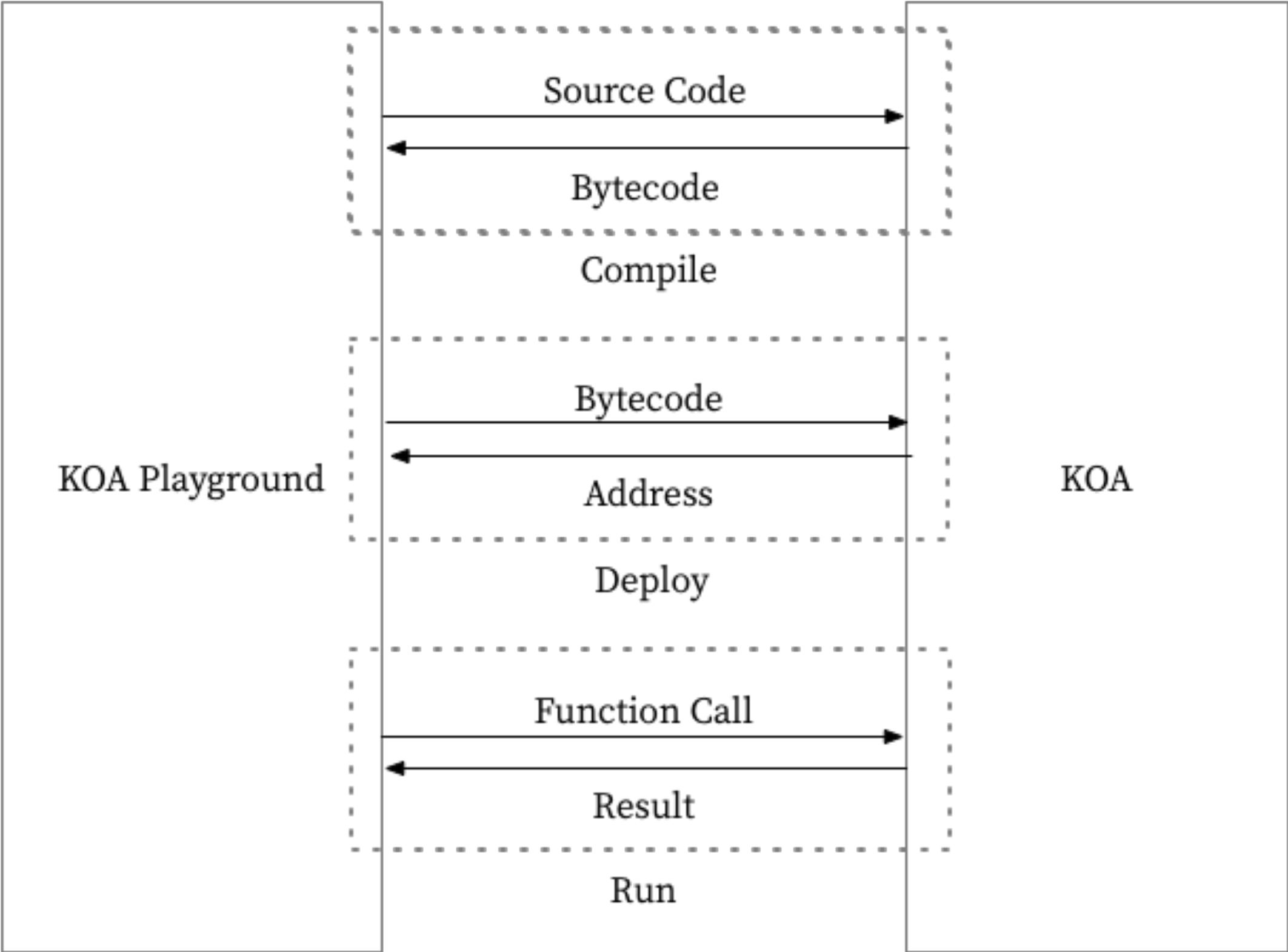}
    \caption{\texttt{koa} system.}
    \label{fig:system}
\end{figure}

\begin{figure}[t]
    \centering
        \includegraphics[width =1.0\linewidth]{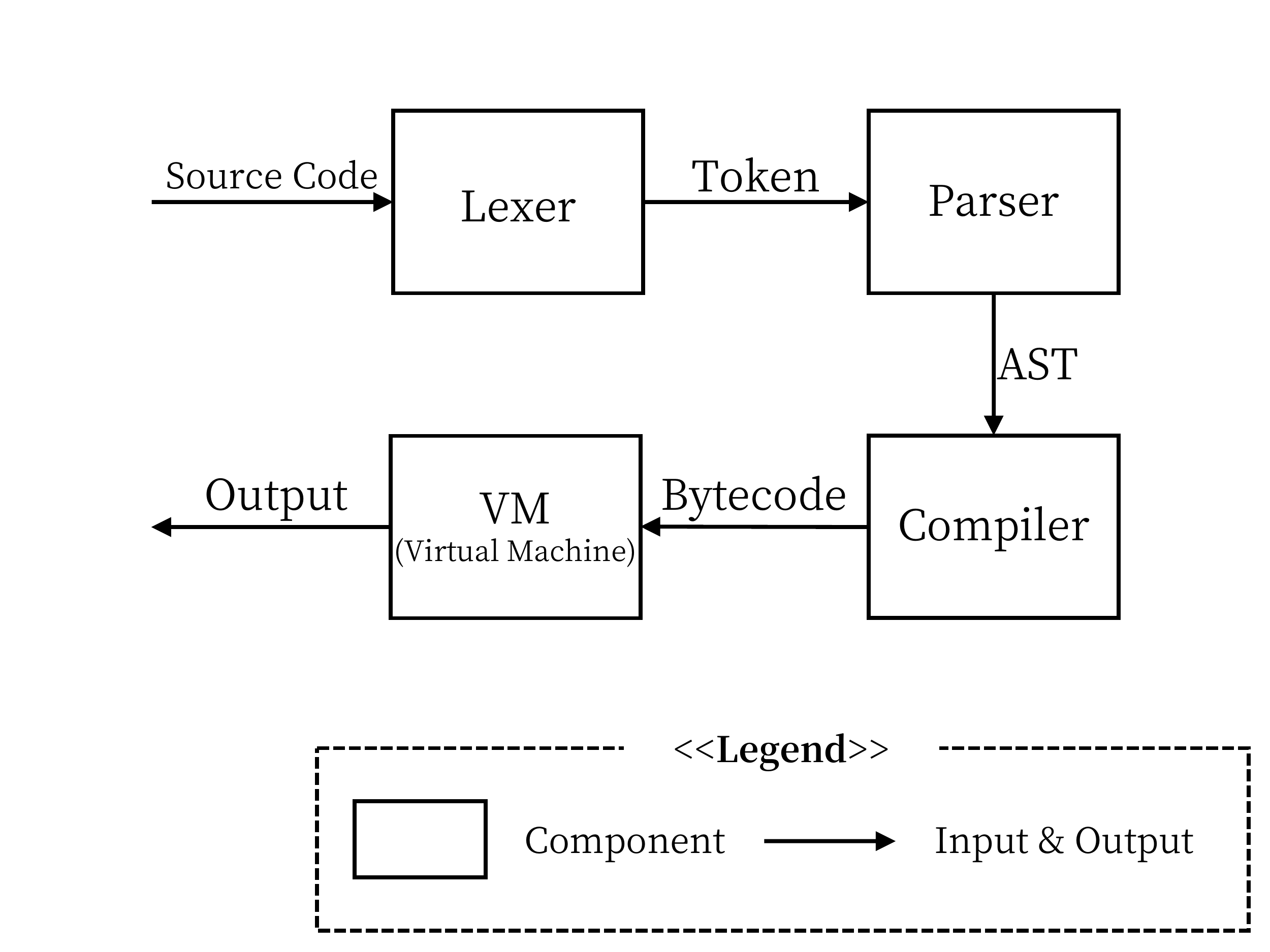}
    \caption{\texttt{koa} architecture.}
    \label{fig:architecture}
\end{figure}

\begin{figure}[t]
    \centering
        \includegraphics[width =1.05\linewidth]{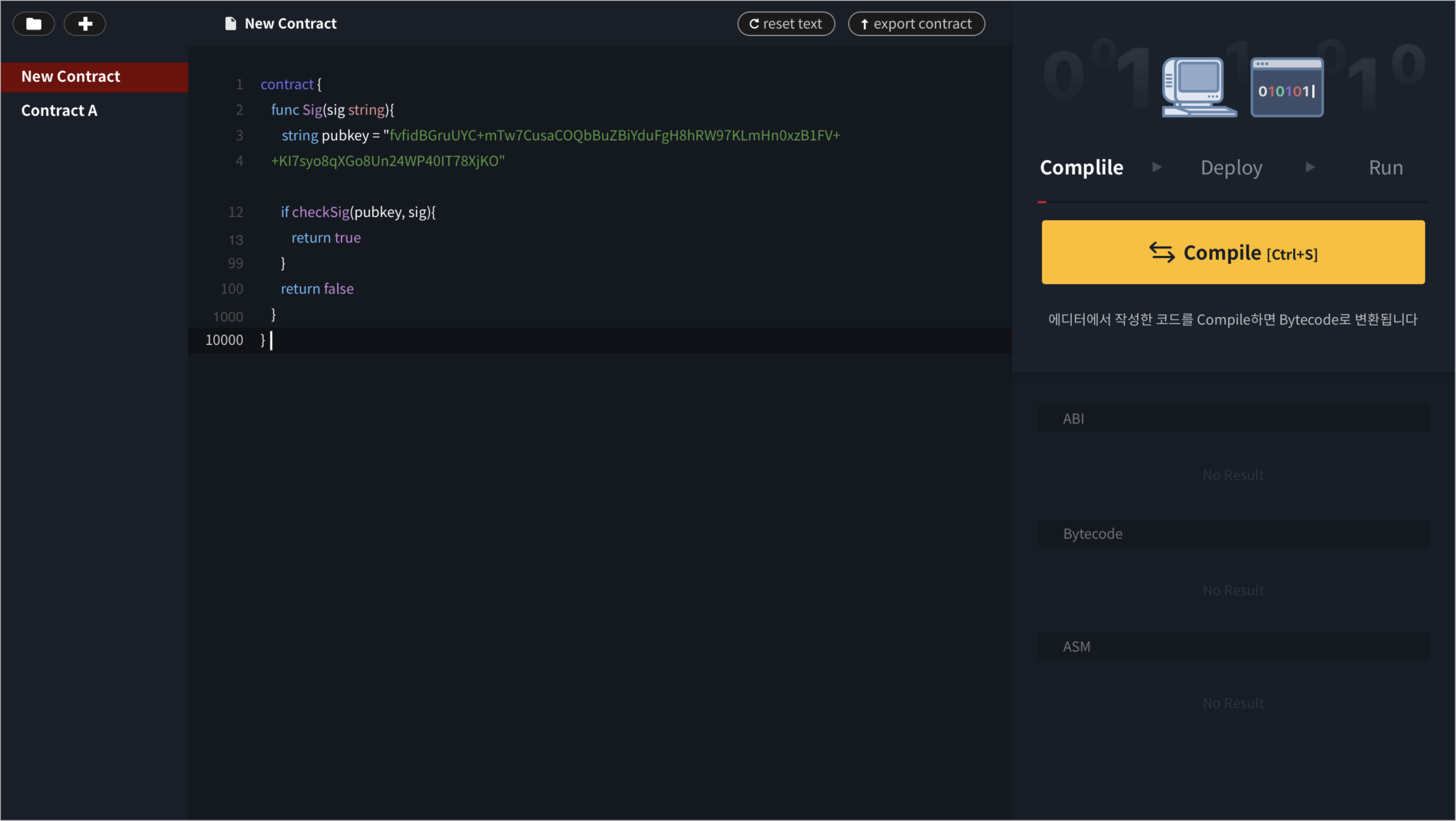}
    \caption{\texttt{koa} playground.}
    \label{fig:playground}
\end{figure}

The structure of the \texttt{Koa} programming language proposed in as Fig.~\ref{fig:system} and it is mainly divided into two parts.
First of all, \texttt{Koa} that performs \textit{lexing}, \textit{parsing}, \textit{compiling}, and \textit{vm} running steps sequentially, as shown in Fig.~\ref{fig:architecture}.
In addition, \texttt{Koa} playground is a kind of IDE environment software platforms and its corresponding web IDE is as presented in Fig.~\ref{fig:playground}.
When users write and compile a source code, it generates a bytecode through lexing, parsing, and compiling steps.
If users deploy this bytecode, it would be registered in blockchain and users receive an address where the program is deployed.
When users conduct execution call to the address for the desired functions, they can get execution results.

\begin{figure}[t]
    \centering
        \includegraphics[width =1.05\linewidth]{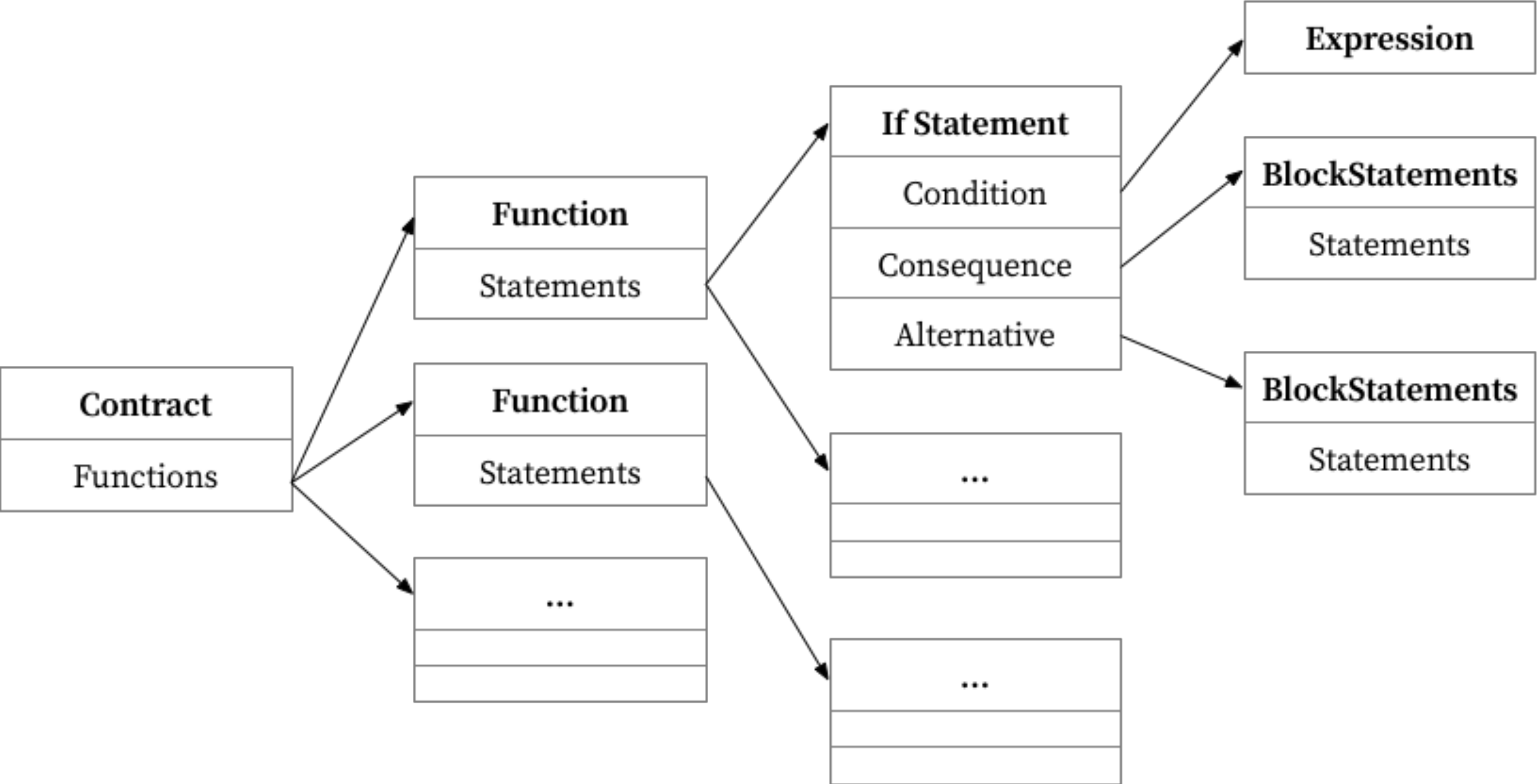}
    \caption{\texttt{koa} AST.}
    \label{fig:ast}
\end{figure}

\section{Implementation}
The form of abstract syntax tree (AST) can be illustrated as Fig.~\ref{fig:ast}.
\texttt{Koa} parser uses \textit{Pratt Parsing} which is recursive parsing which uses the parsing function for each token instead of grammar rules.
It makes the parser intuitive and easy to understand.
Furthermore, \texttt{Koa} receives tokens from the lexer using \textit{TokenBuffer}.
This allows the parser to take responsibility for syntax analysis (parsing) without having tokens directly.

The bytecode generated by \texttt{koa} is passed to Virtual Machine (VM) and then executed.
The VM operations include Stack, Memory, and Heap.
First of all, the stack is used for storing the data and numerical operators.
The, the memory is used to store the data, e.g., variables.
Lastly, the heap is used to store the data which have dynamic sizes, e.g., string.
When users send a call for requesting a function, VM stores this call data in \textit{callfunc}.
Then, it executes the bytecode with stack, memory, and heap.
When all functions are called by users, it returns the output and the program exits.

\begin{acks}
This research was supported by the National Research Foundation of Korea (2016R1C1B1015406, 2017R1A4A1015675); Institute for Information \& Communications Technology Promotion (IITP) grant funded by the Korea government (MSIT) (No. 2018-0-00170, Virtual Presence in Moving Objects through 5G); and IITP grant funded by the Korea government (MSIP) (No. 2017-0-00068, A Development of Driving Decision Engine for Autonomous Driving using Driving Experience Information).
J. Kim is the corresponding author of this paper.
\end{acks}

\bibliographystyle{ACM-Reference-Format}
\bibliography{jhyi16-bib}

\end{document}